\documentclass[fleqn,usenatbib]{mnras}

\usepackage{newtxtext,newtxmath}
\usepackage[T1]{fontenc}
\usepackage{ae,aecompl}
\usepackage{graphicx}	
\usepackage{amsmath}	
\usepackage{amssymb}	
\usepackage{marvosym}
\usepackage{hyperref}

\newcommand{\beq}[1]{\begin{equation}\label{#1}}
\newcommand{\eeq}{\end{equation}}
\newcommand{\beqn}[1]{\begin{eqnarray}\label{#1}}
\newcommand{\eeqn}{\end{eqnarray}}
\newcommand{\sub}[1]{_\mathrm{#1}}

\newcommand{\Op}{\Omega\sub{p}}
\newcommand{\Ocri}{\Omega\sub{c}}
\newcommand{\Rc}{R\sub{c}}
\newcommand{\Mc}{M\sub{c}}
\newcommand{\rhoe}{\rho\sub{e}}
\newcommand{\rhoc}{\rho\sub{c}}
\newcommand{\nm}{n\sub{m}}

\newcommand{\Rp}{R\sub{p}}
\newcommand{\RH}{R\sub{H}}

\newcommand{\Mp}{M\sub{p}}

\newcommand{\Mjup}{\mathrm{M}\sub{Jupiter}}
\newcommand{\Msun}{\mathrm{M}_{\sun}}
\newcommand{\Mearth}{\mathrm{M}_{\oplus}}
\newcommand{\Mmoon}{M\sub{m}}

\newcommand{\Mstar}{M\sub{*}}
\newcommand{\trt}{t\sub{round-trip}}

\newcommand{\amoon}{a\sub{m}}
\newcommand{\apos}{a\sub{p}}
\newcommand{\nmm}{n\sub{m}}

\newcommand{\npp}{n\sub{p}}

\newcommand{\Tors}{\tau\sub{p-*}}
\newcommand{\Torp}{\tau\sub{p-m}}
\newcommand{\Torm}{\tau\sub{m-p}}

\newcommand{\Kpp}{k\sub{2}}
\newcommand{\Qp}{Q}

\newcommand{\Ip}{I\sub{p}}
\newcommand{\Der}{\mathrm{d}}
\newcommand{\AU}{\mathrm{au}}

\newcommand{\quasistatic}{{\it quasistatic}}
\newcommand{\unresponsive}{\textit{unresponsive}}
\newcommand{\dynamic}{\textit{dynamical}}
\newcommand{\realistic}{\textit{realistic}}

\def\sgn{\mathrm{sgn}\,}

\newcommand{\hl}[1]{\textcolor{black}{#1}}
\newcommand{\hll}[1]{\textcolor{black}{#1}}

\title[Migration of exomoons around evolving planets]{The effect of close-in giant planets' evolution on tidal-induced migration of exomoons}

\author[Alvarado-Montes et al.]{
J. A. Alvarado-Montes\thanks{E-mail: \href{mailto:jaimea.alvarado@udea.edu.co}{jaimea.alvarado@udea.edu.co}}, 
Jorge I. Zuluaga and Mario Sucerquia
\\
Solar, Earth and Planetary Physics Group (SEAP)\\ 
Computational Physics and Astrophysics Group (FACom)\\
Instituto de F\'{\i}sica - FCEN, Universidad de Antioquia, Colombia\\ Calle 70 No. 52-21, Medell\'{\i}n, Colombia\\
}

\date{Accepted 2017 July 8. Received 2017 July 8; in original form 2017 April 6}
\pubyear{2017}

\begin{document}
\label{firstpage}
\pagerange{\pageref{firstpage}--\pageref{lastpage}}
\maketitle

\begin{abstract}

\hl{Hypothetical exomoons around close-in giant planets may migrate inwards and/or outwards in virtue of the interplay of the star, planet and moon tidal interactions.  These processes could be responsible for the disruption of lunar systems, the collision of moons with planets or could provide a mechanism for the formation of exorings. Several models have been developed to determine the fate of exomoons when subject to the tidal effects of their host planet. None of them have taken into account the key role that planetary evolution could play in this process. In this paper we put together numerical models of exomoon tidal-induced orbital evolution, results of planetary evolution and interior structure models, to study the final fate of exomoons around evolving close-in gas giants. We have found that planetary evolution significantly affects not only the time-scale of exomoon migration but also its final fate.  Thus, if any change in planetary radius, internal mass distribution and rotation occurs in time-scales lower or comparable to orbital evolution, exomoon may only migrate outwards and prevent tidal disruption or a collision with the planet. If exomoons are discovered in the future around close-in giant planets, our results may contribute to constraint planetary evolution and internal structure models.}
\end{abstract}


\section{Introduction}
\label{sec:introduction}

\begin{figure*}
    \centering
        \includegraphics[scale=0.4]{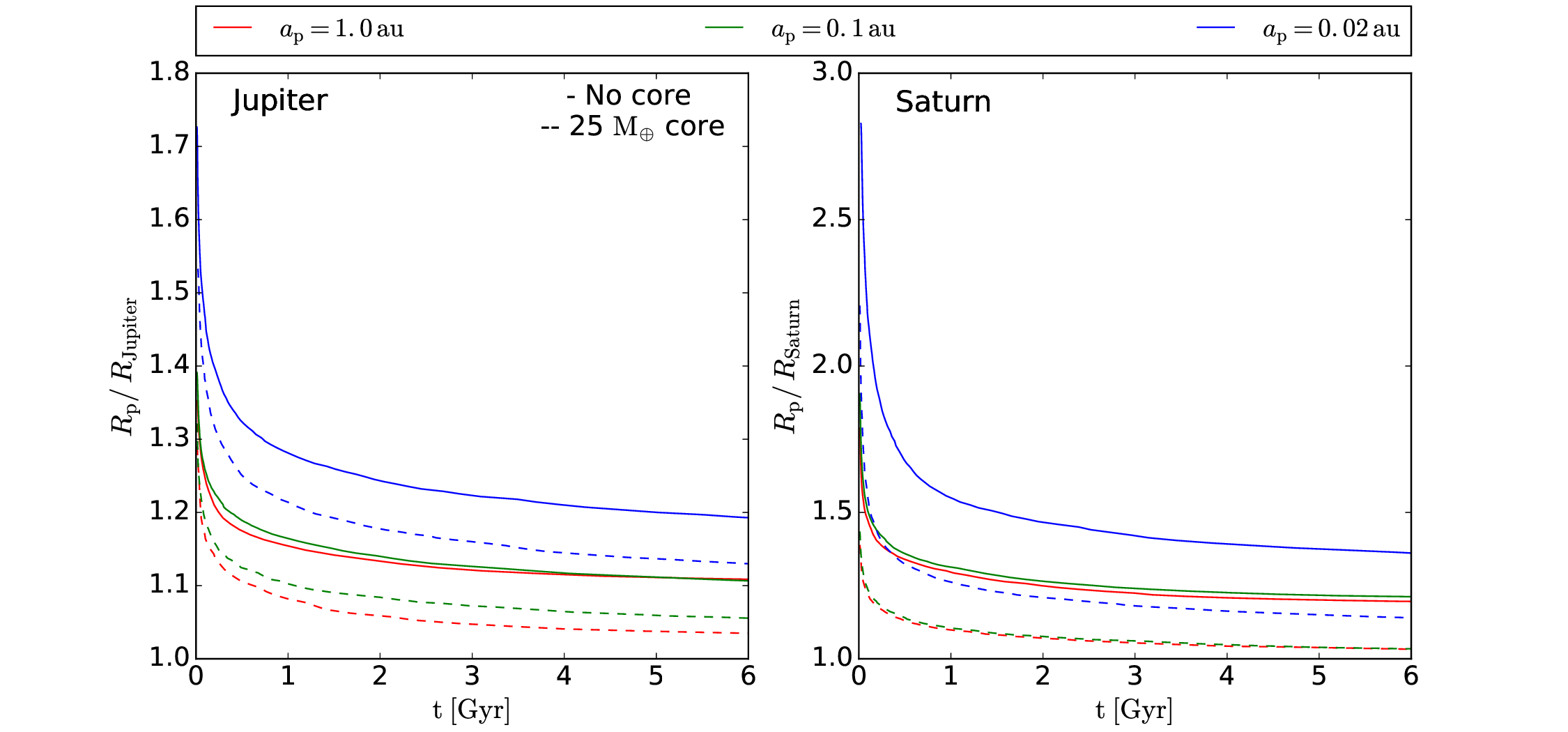}
    \caption{Evolution of planetary radius for two sets of planets: Jupiter-like (left panel) and Saturn-like (right panel) planets \citep{Fortney2007}. Solid and dashed lines correspond to planets with no solid core and with a solid core of 25 $\Mearth$. Evolution has been computed at different distances from the host star: 0.02 (blue), 0.1 (green) and 1.0 $\AU$ (red).}
\label{fig:radiusEvolution}
\end{figure*}

Since the discovery of the first exoplanet \citep{Mayor1995} it has been clear that migration of giant planets is a common phenomenon in planetary systems (see e.g. \citealt{Armitage2010} and references therein). The question if those migrated planets have moons, i.e. if their putative moons survive migration and/or the formation process succeeds in those conditions, remains still open \citep{Barnes2002,Gong2013,Heller2015}. \hl{Additionally,} despite multiple efforts intended to \hl{discover} the first exomoon (see e.g. \citealt{Kipping2013,Heller2014}) none has been detected among the hundreds of giant planets discovered up to now. 

\hl{If exomoons survive planetary migration}, their gravitational and tidal interaction with \hl{their host} planet, other \hl{bodies in the planetary system} and the star, will modify their original orbits \citep{Barnes2002}. \hl{Several} authors have studied the dynamics of exomoons around close-in giant planets \citep{Barnes2002,Sasaki2012,Gong2013} \hl{systematically finding} that exomoons \hl{may} migrate inwards and/or outwards \hl{around the planet} and \hl{their final fate} could be diverse. \hl{Depending on their mass and initial orbit, migrating exomoons may}: \hl{collide with the planet}, \hl{be disrupted or obliterated inside Roche radius \citep{Esposito2002,Charnoz2009,Bromley2013}},  be ejected from the planetary system, or become a new \hl{(dwarf)} planet (a `ploonet'). 

Most of the theoretical models \hl{used to describe} exomoon migration, apply well-known analytical formalisms that have been tested in the solar system and beyond. Others use numerical simulations taking into account complex and analytically intractable aspects of the problem \citep{Namouni2010,Sasaki2012}. All of them\hl{, however,} have one feature in common: exomoon interactions and migration happen while the planet remains unchanged. 

As opposed to solid planets, gas giants \hl{may} change significantly during the first hundreds of Myrs (see \citealt{Fortney2007} and references therein). In particular, planetary radius and \hl{tidal-related properties} (gyration radius, love number, \hl{dissipation reservoir}, etc.) may change \hl{in relatively short time-scales}. \hl{The change in these properties, that among other effects, determines the strength of the tidal interaction between the star, the planet and the moon, may also} have an important effect in the orbital evolution of exomoons.

In this work we explore how the final fate of an exomoon can be affected by the evolution of the physical properties of its host planet. We \hl{aim at computing  the time-scale of exomoon orbital evolution taking into account the change in planetary properties that may take place during migration}. For this purpose, we use the same basic analytical and numerical formalism applied in previous works \citep{Barnes2002,Sasaki2012}, but \hl{plug into them the results of planetary evolution models \citep{Fortney2007} and recent analytical formulas developed to calculate the tidal-related properties of fluid giant planets \citep{Ogilvie2013,Guenel2014,Mathis2015a}}.  

This paper \hl{is organized} as follows:  In Section \ref{sec:evolution} \hl{we present and describe the results of the planetary evolution model we adopted here, and the analytical formulas we used to describe the tidal-related properties of a bi-layer fluid planet}. \hl{Section \ref{sec:migration} explains the physics of 
exomoon migration and the formalism used to describe it.  In Section \ref{sec:results} we apply our model to study exomoon migration under different evolutionary scenarios. Finally, Section \ref{sec:discon} summarizes our results and discusses the limitations, implications and future prospects of our model.}

\section{\hl{The evolution of gas giants}}
\label{sec:evolution}

\hl{Models of thermal evolution and internal structure of giant planets have been available in literature for several decades. Furthermore, in the last 10 years improved and more general models have been developed to describe the evolution of planets in extreme conditions, given that many of them have been discovered in those environments} \citep{Guillot2006,Fortney2007,Garaud2007}.  \hl{To illustrate  and test our models of exomoon-migration under the effect of an evolving planet}, we will restrict to the \hl{well-known} results published by \citet{Fortney2007}. \hl{These results encompass a wide range of planetary masses, compositions and distances to the host star}.

\hl{In \autoref{fig:radiusEvolution} we show the evolution of radius for two families of planets: Jupiter and Saturn analogues, with different compositions (core mass) and located at different distances from a solar-mass star. The time-scale of planetary radius evolution strongly depends on planetary mass and distance to the star}. At close distances, namely $a_P\sim 0.1\;\AU$, a Jupiter-like planet will change its radius by $\sim 30$ per cent within the first Gyr, while a Saturn-analogue will shrink almost by a factor of 2.5 ($\sim 60$ per cent) in the same time.  \hl{On the other hand, although} planetary radius at a given time strongly depends on composition, the  fractional evolution as well as the time-scale \hl{remain the same for planets with different core mass}.

\hl{Even though planetary radius and evolutionary time-scales depend on the distance to the star, and therefore we should model the evolution of our planets depending on their assumed orbital radius,} we notice that for masses between 0.3 and 3 $\Mjup$, planets with the same composition, have nearly the same radii\hl{, provided their distances are in the range $a_{\rm P}=0.1-1.0\;\AU$} (see Fig. 3 of \citealt{Fortney2007}). \hl{Planets much closer than 0.1 $\AU$ are considerably larger}. \hl{Hereafter,} we will use \hl{in all our models, the properties calculated for planets at 1 $\AU$}. However, this implies that in our results, especially for planets migrating at distances $\ll1\;\AU$, planetary radii and densities could be \hl{slightly} under and overestimated, respectively. 

\hl{A young gas planet, remains `inflated' for up to a billion of years. The evolution of its radius comes mostly from the evolution of its extended atmosphere.  Its deep interior, however, does not change significantly in the same time-scale neither in composition, nor in mass or radius. Hereafter we will assume that the mass and radius of the planetary solid core remain constant during the relevant time-scales studied in this work.}

A change in planetary radius, mean density and rotational rate, would have a significant impact in \hl{other key mechanical and gravitational properties of the planet.  For the purposes pursued here we will focus on the evolution of the so-called ``tidal dissipation reservoir'' \citep{Ogilvie2013}, which in the classical tidal theory is identified with the ratio $\Kpp/\Qp$; being $k_2$ the  (frequency-independent) love number and $\Qp$ the so-called ``tidal dissipation quality factor'' (the ratio of the energy in the equilibrium tide and the energy dissipated per rotational period).}

\begin{center}
\begin{figure}
\includegraphics [scale=0.42]{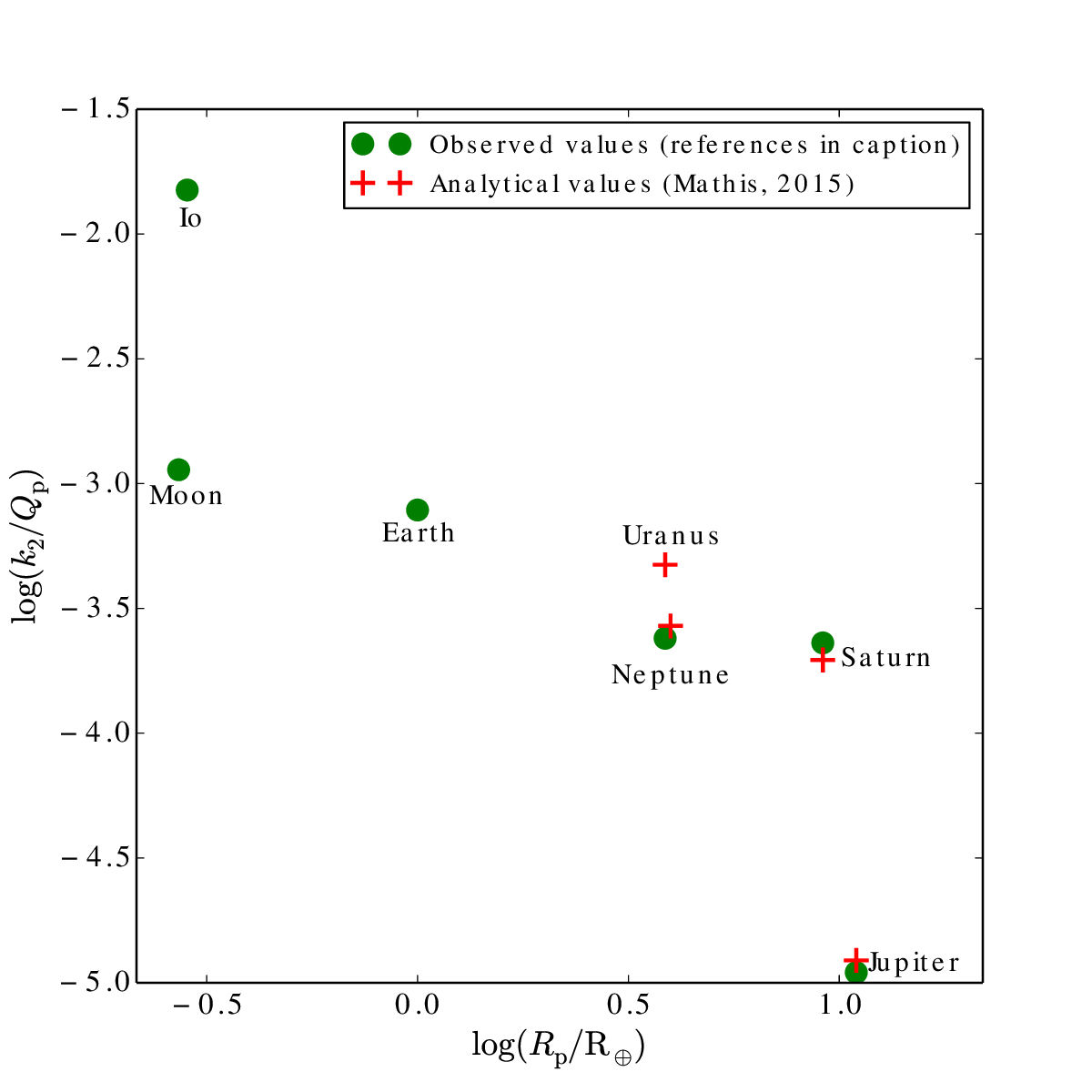}
\caption{\hl{Values of $\Kpp/\Qp$ for several Solar System bodies: Jupiter and Io \citep{Lainey2009}; Saturn \citep{Lainey2012}; the Moon \citep{Dickey1994}; the Earth \citep*{Ray1996,Kozai1968} and Neptune \citep{Trafton1974}. The tidal dissipation reservoir is larger in solid bodies (Io, Moon and Earth) than in planets having extensive gaseous envelopes (Neptune, Saturn and Jupiter).  Red crosses show the values of $\Kpp/\Qp$ estimated with the analytical formula in Eq. (\ref{eq:k2QFormula}), assuming values for $\alpha$, $\beta$ and $\gamma$} as given by \citet{Mathis2015a}.}
\label{fig:k2Qvalues}
\end{figure}
\end{center}

\hl{Usually, $\Kpp$ and $\Qp$ are estimated from observations of actual bodies (planets and stars).  In \autoref{fig:k2Qvalues} we present the measured value of $\Kpp/\Qp$ for several Solar System objects. In the case of exoplanets, $\Kpp/\Qp$ is estimated using simplifying assumptions about their interior structures, which  are usually assumed static ($\Kpp/\Qp$ constant). For our purpose here, we need to estimate theoretically the value of $\Kpp/\Qp$ as a function of the evolving bulk and interior properties of the planet.}

\hl{Love numbers, $k_l^m(\omega)$, are, in general, complex frequency-dependent coefficients that measure the response of the planet to tidal stresses. Under simplifying assumptions (low planetary obliquity and moon orbit-planetary axis relative inclination), the only non-negligible love number is $\Kpp^2(\omega)$. In physical terms $\Kpp^2(\omega)$ provide information about the dissipation of tidal energy inside the planet and the interchange of rotational and orbital angular momentum among the planet and the moon.}

\hl{In particular the frequency averaged imaginary part of  $\Kpp^2(\omega)$ provides an estimation of $\Kpp/\Qp$,}

\beq{eq:Imk2}
\Kpp/\Qp=\int_{-\infty}^{+\infty}{\rm Im}[\Kpp(\omega)]\frac{d\omega}{\omega}=\int_{-\infty}^{+\infty}\frac{|\Kpp^2(\omega)|}{Q_2^2(\omega)}\frac{d\omega}{\omega}
\eeq

\hl{In recent years, several advances have been achieved in the understanding and description of tidal dissipation in the fluid interiors of gas giants \citep{Ogilvie2007,Ogilvie2013,Guenel2014,Mathis2015a}. In particular, analytical expressions for the tidal dissipation reservoir as a function of the bulk properties of simplified bi-layer fluid planets have been obtained \citep{Ogilvie2013}. For our investigation we will apply the formula adapted by \citet{Guenel2014} from  \citet{Ogilvie2013}:}

\beq{eq:k2QFormula}
\frac{\Kpp}{\Qp} = \frac{100\pi}{63}\epsilon^{2}\frac{\alpha^{5}}{1-\alpha^{5}}\left[1+
	\frac{1-\gamma}{\gamma}\alpha^{3}\right]\left[1+
		\frac{5}{2}\frac{1-\gamma}{\gamma}\alpha^{3}\right]^{-2}
\eeq 

Here $\epsilon^{2}\equiv (\Omega/\Ocri)^{2}$, where $\Omega$ is the planetary rotational rate and    $\Ocri=(G\Mp/\Rp^{3})^{1/2}$ is the critical rotation rate; $\alpha$, $\beta$ and $\gamma$ are dimensionless parameters defined \hl{in terms of the bulk properties of the planet} as:

\beq{eq:k2Qparameters}
\alpha=\frac{\Rc}{\Rp},\quad \beta=\frac{\Mc}{\Mp}\quad\text{and}\quad \gamma=\frac{\rhoe}{\rhoc}=
	\frac{\alpha^{3}(1-\beta)}{\beta(1-\alpha^{3})},
\eeq

\noindent 
where $\Rc\,(\Rp)$, $\Mc\,(\Mp)$ and $\rhoc$ are the radius, the mass and the density of the core, respectively; $\rhoe$ is the density of the (fluid) envelope.

\hl{This formula is valid only under very specific conditions.  We are assuming that the planet is made of two uniform layers: a fluid external one and a fluid or solid denser core.  In our models tidal dissipation occurs only in the turbulent fluid envelope; for the sake of simplicity we are neglecting the inelastic tidal dissipation in the core \citep{Guenel2014}. Dissipation in the fluid layer arises from turbulent friction of coriolis-driven inertial waves (this is the reason of the strong dependence of $\Kpp/\Qp$ on the rotational rate $\epsilon$). The approximation in Eq. (\ref{eq:k2QFormula}) also breaks down if centrifugal forces are significant and therefore it only applies in the case when the planet rotates very slowly, i.e. when $\epsilon\ll 1$.}

\hl{\citet{Mathis2015a} estimated the values of $\alpha$ and $\beta$ for Jupiter, Saturn, Uranus and Neptune that better match the measured values of $\Kpp/\Qp$ (see Table 1 in \citealt{Mathis2015a} and red crosses in \autoref{fig:k2Qvalues} here). To that end, he assumed educated estimations of the unknown internal bulk properties of the Solar System giants, namely $\Rc, \Mc$ and $\rhoc$.}

\hl{In order to apply \autoref{eq:k2QFormula} to the purposes pursued here, we need to provide an estimate to the functions of $\alpha(t)$, $\beta(t)$ and $\epsilon(t)$ or equivalently $\Rc(t)$, $\Rp(t)$, $\Mc(t)$ and $\Op(t)$.  For simplicity, we will assume that the liquid/solid core of the planet is already formed or evolves very slowly during most of the envelope contraction, i.e. $\beta(t)=\beta(0)$.  $\epsilon(t)$ will be computed consistently from the radius provided by the planetary evolution models (\autoref{fig:radiusEvolution}) and the instantaneous rotational rate of the planet, which changes due to the tidal interaction with the star and the moon.  For $\alpha(t)$, we will assume an asymptotic value $\alpha(\infty)$, similar to that of Solar System planets (Table 1 in \citealt{Mathis2015a}); the value at any time will be obtained from:}

\beq{eq:alpha}
\alpha(t)=\alpha(\infty) \frac{R_p(\infty)}{R_p(t)}
\eeq

\begin{center}
\begin{figure}
\includegraphics[scale=0.45]{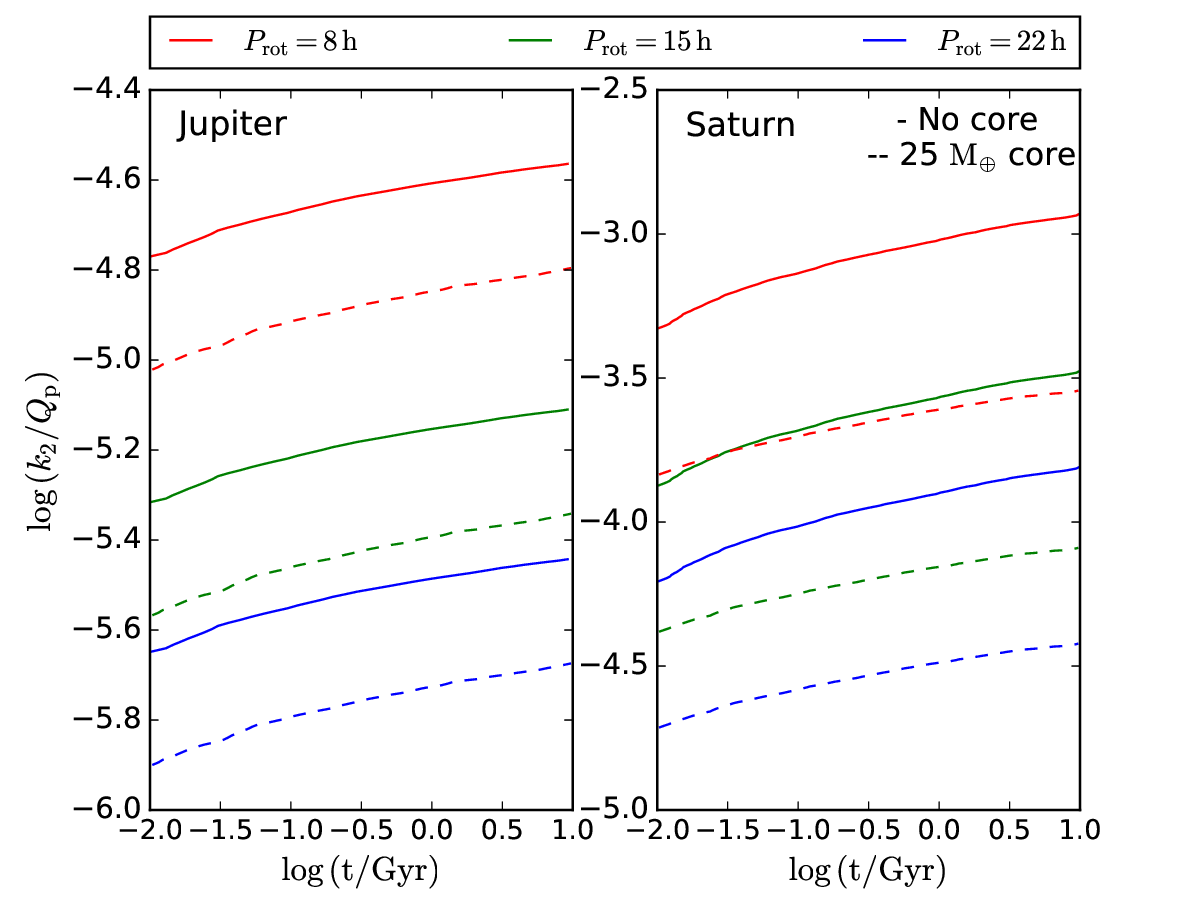}
\caption{\hl{Evolution of the tidal dissipation reservoir $\Kpp/\Qp$ for Jupiter and Saturn analogues orbiting a Solar-mass star at 1 au.  Colors correspond to different rotational periods (shown at the top of the graph). Solid and dashed lines correspond to planets with no core ($\beta\to\infty$) and a liquid/solid core of 25$\Mearth$, respectively.}}
\label{fig:k2Qevolution}
\end{figure}
\end{center}

\hl{In \autoref{fig:k2Qevolution} we show the result of applying Eq. (\ref{eq:k2QFormula}) and the previously described prescription, to compute the evolution of $\Kpp/\Qp$ for Jupiter and Saturn analogues at 1 AU. For illustration purposes, a set of different constant rotational rates is assumed.  In both cases we have assumed asymptotic $\alpha$ values equal to that of Jupiter and Saturn respectively.} 

\hl{As expected $\Kpp/\Qp$ strongly depends on rotational rate.  A change of $P\sub{rot}$ by a factor of 3, change $\Kpp/\Qp$ by almost one order of magnitude.  As the planet contracts the value of $\alpha$ goes up as $1/\Rp$; at the same time $\epsilon^2$ goes down as $\Rp^3$. For small values of $\alpha$,  $\Kpp/\Qp\propto\epsilon^2\alpha^5\propto\Rp^2$, the planet's contraction produces a net increase in tidal dissipation.}

\section{Moon orbital migration}
\label{sec:migration}

\hl{Let us now consider an already formed close-in giant planet with mass $\Mp$, initial radius $\Rp(0)$ and rotation rate $\Op(0)$, around a star of mass $\Mstar$.  It is also assumed that the planet is in a final nearly circular orbit with semi major axis $a\sub{p}$ (orbital period $P\sub{orb}$). In addition, we suppose that the planet has migrated from its formation place to its final orbit, in a time scale much shorter than its thermal evolution \citep{Tanaka2002,Bate2003,Papaloizou2006,Armitage2010}. At its final orbit the planet harbors a regular moon (orbital motion in the same direction as planetary rotation) with mass $M\sub{m}$, and whose formation process and/or mechanism to survive migration are irrelevant here. The moon is orbiting the planet at an initial nearly circular orbit, with a semi major axis parametrized as $\amoon(0)=f\Rp(0)$.} Thus, for instance $f \simeq 4$
for Saturn's moon Enceladus and $f \simeq 10$ for Jupiter's moon
Europa.

\hl{The star-planet-moon interaction raises a complex tidal bulge on the planet, which gives rise to star-planet and moon-planet torques, that according to the \hll{constant time-lag model} are given by \citep{Murray2000}:}

\beq{eq:Tors}
\Tors  \approx  -\frac{3}{2}\frac{\Kpp}{\Qp}\frac{ G \Mstar^{2} \Rp^{5}}{
  \apos^{6}}\,\sgn(\Op-\npp)
\eeq

\beq{eq:Torp}
\Torp  \approx  -\frac{3}{2}\frac{\Kpp}{\Qp}\frac{G \Mmoon^{2} \Rp^{5}}{
  \amoon^{6}}\,\sgn(\Op-\nmm)
\eeq

\hll{The latter consideration does not lead to discontinuities for vanishing tidal frequencies (e.g. synchronous rotation), and it sheds light on a thorough analytical assessment of the tidal-related effects, free from assumptions on the eccentricity (which is crucial when it comes to close-in exoplanets, see \citealt{Leconte2010}). Still, the constant time-lag model is a linear theory and taking into account nonlinear terms can result in noticeable changes \citep{Weinberg2012}.}

\begin{figure*}
{ \centering 
\includegraphics [width=130mm]
{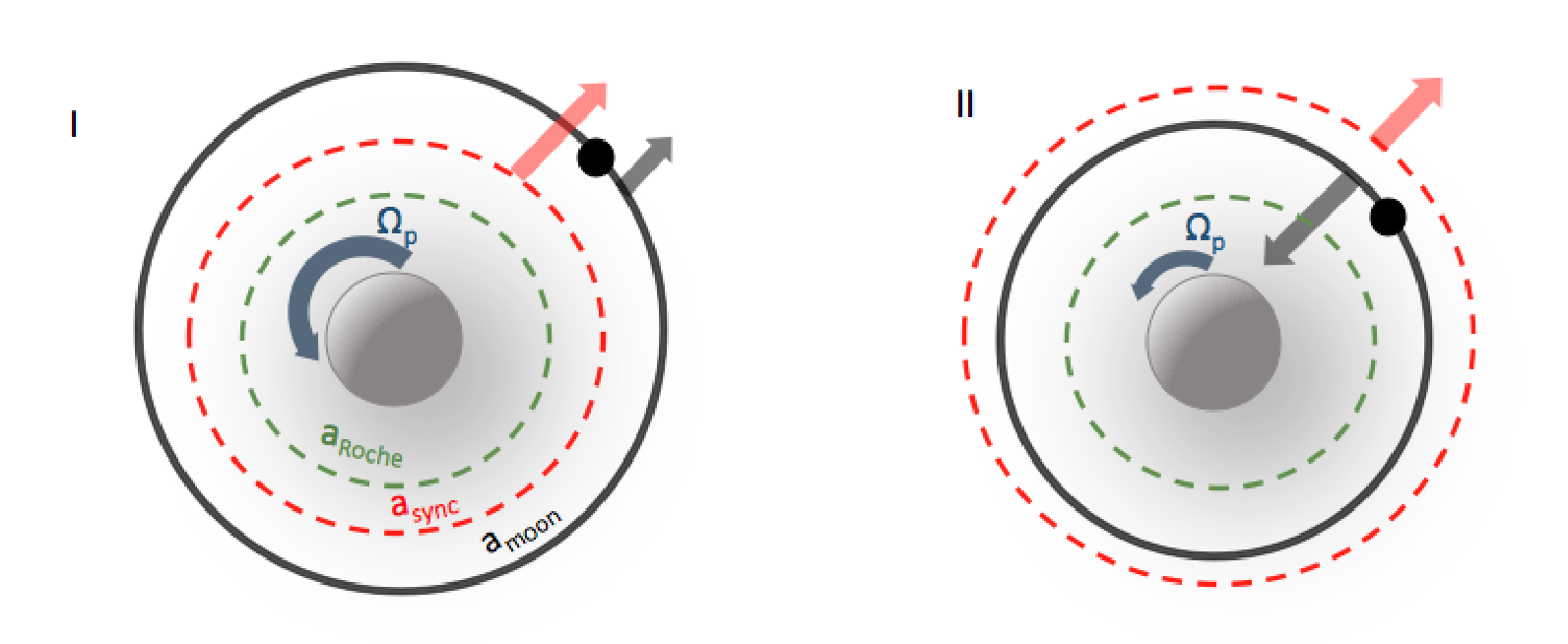}
\includegraphics [width=130mm]
{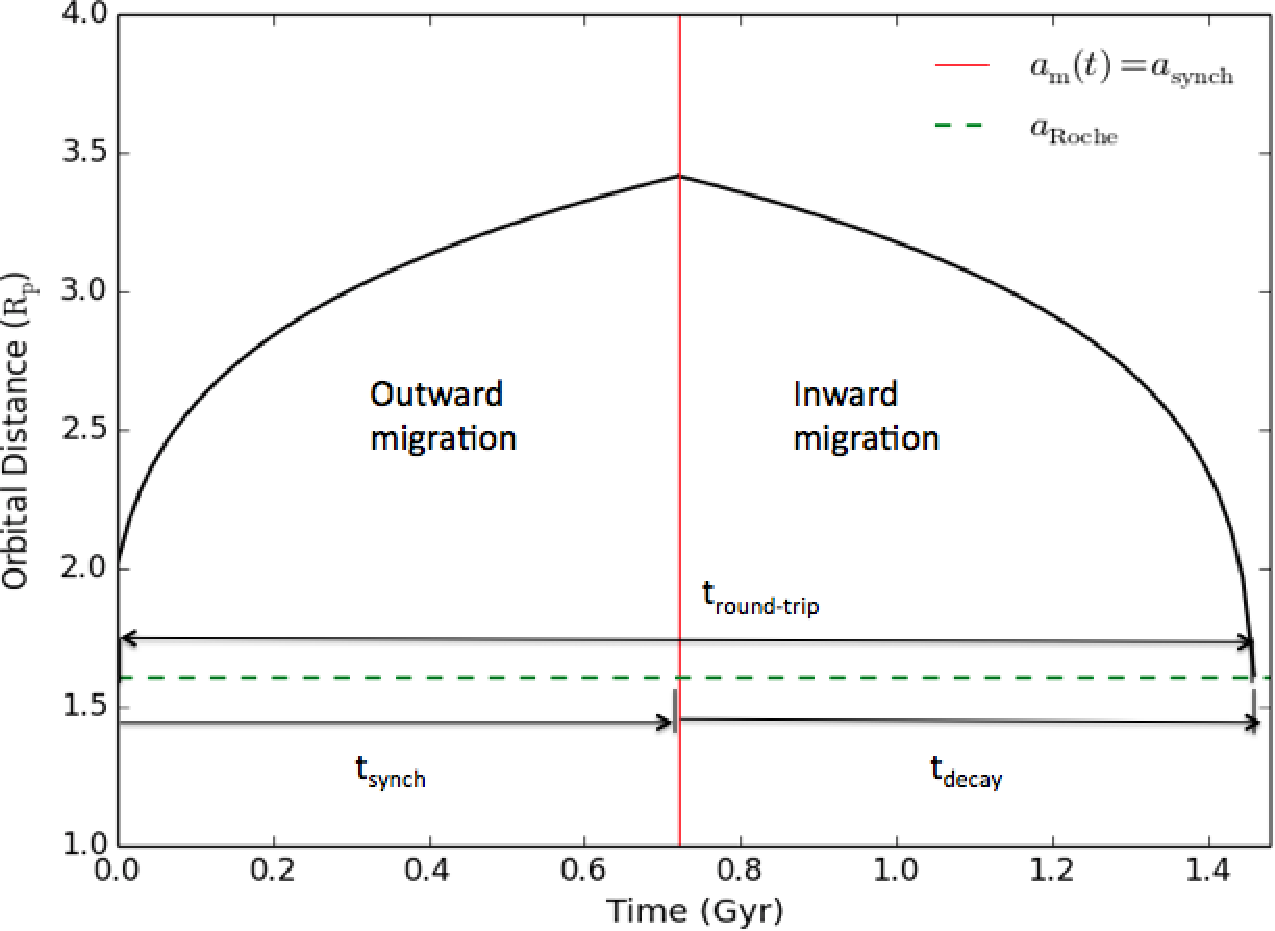}
\caption{\hl{Moon orbital migration under the interplay of the star-planet-moon tidal interactions.  
{\it Upper panel:} a schematic representation of the process.  
{\it Lower panel:} an actual integration of the tidal evolution differential equations, for a Saturn analogue, located at an orbit with $P\sub{orb}=80$ days around a Solar-mass star.  The moon is an Enceladus-mass object with an initial orbit of $\amoon(0)=2 \Rp(0)$.  For obtaining this plot we assumed that the values of $\Kpp/\Qp$ and $\Rp$ remain constant (\quasistatic\,scenario).}}
\label{fig:migrationSchematic}}
\end{figure*}

\hl{In the linear regime the torques exerted by the moon and the star add up, $\tau\sub{tot}\approx\Tors+\Torp$, and produce a change in the rotational angular momentum of the planet $L\sub{p}(t)=I\sub{p}(t)\Op(t)$, where $I\sub{p}(t)=\kappa^2\Mp\Rp(t)^2$ is the planet's moment of inertia, and $\kappa^2$ is the gyration radius, given by the Newton second law:}

\beq{eq:2ndlawPlanet}
\frac{dL\sub{p}}{dt}\approx\Tors+\Torp
\eeq

\hl{The reaction to the moon-planet torque, $\Torm=-\Torp$ modifies the orbital angular momentum of the moon $L\sub{m}=M\sub{m}\amoon(t)\nm(t)$ as follows:}

\beq{eq:2ndlawMoon}
\frac{dL\sub{m}}{dt}\approx-\Torp
\eeq

\hl{We are assuming that the orbit of the moon is not modified by stellar perturbations nor the tidal interaction of the moon with the planet.  Although both cases may be relevant in realistic scenarios \citep{Sasaki2012}, the inclusion of these effects will not modify considerably our general conclusions.}

\hl{Under the assumption of a relatively slow planetary radius variation, i.e. $d\Ip/dt\ll (\Ip/\Op) d\Op/dt$, Eqs. (\ref{eq:2ndlawPlanet},\ref{eq:2ndlawMoon}) can be rewritten as:}

\beq{eq:dOdt}
\begin{split}
\frac{\Der\Op}{\Der t}=-\frac{3}{2}\frac{\Kpp}{\Qp}\frac{\Rp^{3}}{k^{2}G}
\left[
\frac{(G\Mstar)^{2}}{\Mp\apos^{6}}\,\sgn(\Op-\npp)+\right.\\
\left.\frac{\Mmoon^{2}}{\Mp^{3}}\nmm^{4}\,\sgn(\Op-\nmm) \right]
\end{split}
\eeq

\beq{eq:dnmdt}
\frac{\Der \nmm}{\Der t}=-\frac{9}{2}\frac{\Kpp}{\Qp}\frac{\Mmoon\Rp^{5}}{ G^{5/3}\Mp^{8/3}}\nmm^{16/3}\sgn(\Op-\nmm).
\eeq

\hl{Under general circumstances, $\Op>\npp$ and $\Mp/\Mstar\gg\Mmoon/\Mp$, and hence $d\Op/dt<0$. Consequently the planet rotation slows down monotonically until it finally ``locks-down'' at around $\Op\sim\npp$.}\footnote{\hl{its final rotational state actually depends, among other factors, on its orbital eccentricity and planetary interior properties, see e.g. \citealt{Cuartas2016}).}}

\hl{On the other hand, the evolution of $\nmm$ is more complex. If the moon starts at a relatively large distance from the planet where $\nmm(0)<\Op(0)$, then $d\nmm/dt<0$.  In addition, the orbital mean motion will slow down and the moon will migrate outwards (see \autoref{fig:migrationSchematic}). However, since the planetary rotational rate is also slowing down, under certain circumstances and at a finite time $t\sub{sync}$, $\nmm(t\sub{sync})=\Op(t\sub{sync})$. At this stage, migration will stop momentarily until $\nmm$ starts to outpace $\Op$ ($\nmm>\Op$, $d\nmm/dt>0$). Therefore, inward migration is triggered and the moon will eventually collide, or be obliterated by the planet inside the Roche radius.}

\hl{We show in \autoref{fig:migrationSchematic} an schematic representation of this process and an actual solution to Eqs. (\ref{eq:dOdt},\ref{eq:dnmdt}) for a typical planet-moon-star system.}

\hl{The described tidal-induced moon migration process has three relevant time-scales:} 

\begin{itemize}

\item \hl{The synchronization time, $t\sub{sync}$ which can be defined by the condition}:

\beq{eq:synch}
\nmm(t\sub{sync})=\Op(t\sub{sync})
\eeq

\item \hl{The decay time, $t\sub{decay}$ given by}:

\beq{eq:decay}
\nmm(t\sub{sync}+t\sub{decay})=n\sub{Roche}
\eeq

\hl{Here $n\sub{Roche}=\sqrt{G\Mp}a\sub{Roche}^{-3/2}$ is the orbital mean motion at the Roche radius.  For the purpose of this work we will assume $a\sub{Roche}\approx 2(\rho\sub{m}/\rho\sub{p})^{1/3}\Rp$ \citep{Roche1849,Esposito2002}, where $\rho\sub{m}$ is the mean density of the moon.}
\item \hl{and the ``round-trip'' time, $t\sub{round-trip}$}:

\beq{eq:roundt}
t\sub{round-trip}=t\sub{sync}+t\sub{decay}
\eeq
\end{itemize}

\section{Results}
\label{sec:results}

\hl{It is usual that when solving Eqs. (\ref{eq:dOdt},\ref{eq:dnmdt}) all the properties in the right-hand side, including the tidal dissipation reservoir $\Kpp/\Op$, planetary radius $\Rp$ and gyration radius $\kappa^2$ (moment of inertia), be assumed as constants.  This is the way the problem has been analyzed in literature \citep{Barnes2002,Sasaki2012}.  In a more realistic case we must allow that these quantities evolve independently, or coupled with orbital migration.  However, in order to individualize the effect that each property has on moon orbital evolution}, we will consider four different evolutionary scenarios we have labeled as \hl{\quasistatic, \unresponsive, \dynamic\ and \realistic}.

\subsection{Quasistatic migration}

In this case \hl{all of the relevant mechanical properties ($\Kpp/\Op$, $\Rp$ and $\kappa$)} are assumed \hl{nearly} constant or \hl{varying very slowly} during moon migration. This \hl{is the most common scenario} found in literature \citep{Barnes2002,Sasaki2012}.  \hl{Although unrealistic, the results obtained in this scenario provide us with first order estimations of the time-scales of moon migration and its dependence on key properties of the system.}

\begin{figure}
\centering
\includegraphics[scale=0.462]{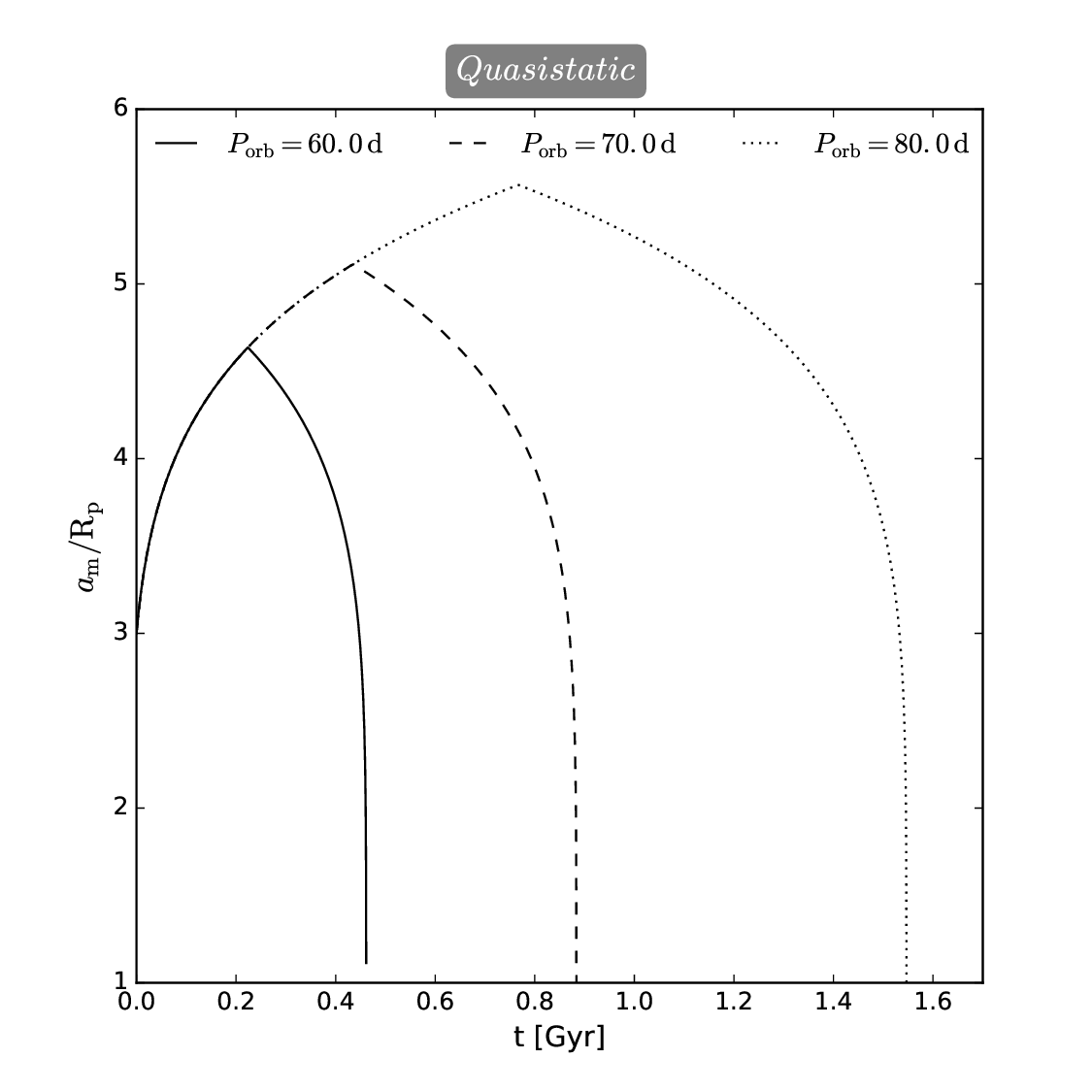}
\caption{Moon orbital semi major axis as function of time for a warm Saturn-Superenceladus system. Solid, dashed and dotted lines correspond to planetary orbital periods of 60, 70 and 80 days, respectively.}
\label{fig:quasistaticEvolution}
\end{figure}

\begin{figure*}
\centering
\includegraphics[scale=0.45]{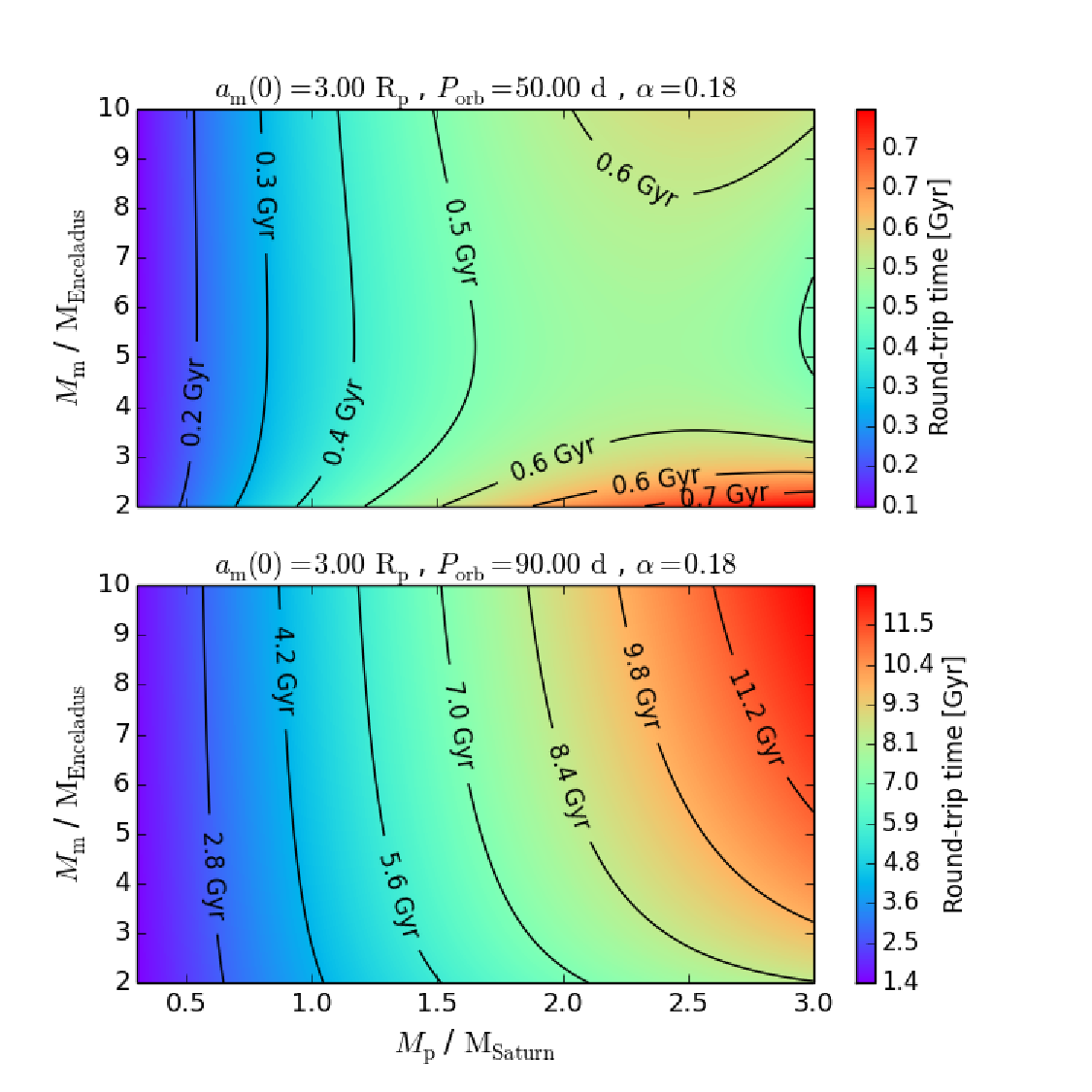}
\includegraphics[scale=0.45]{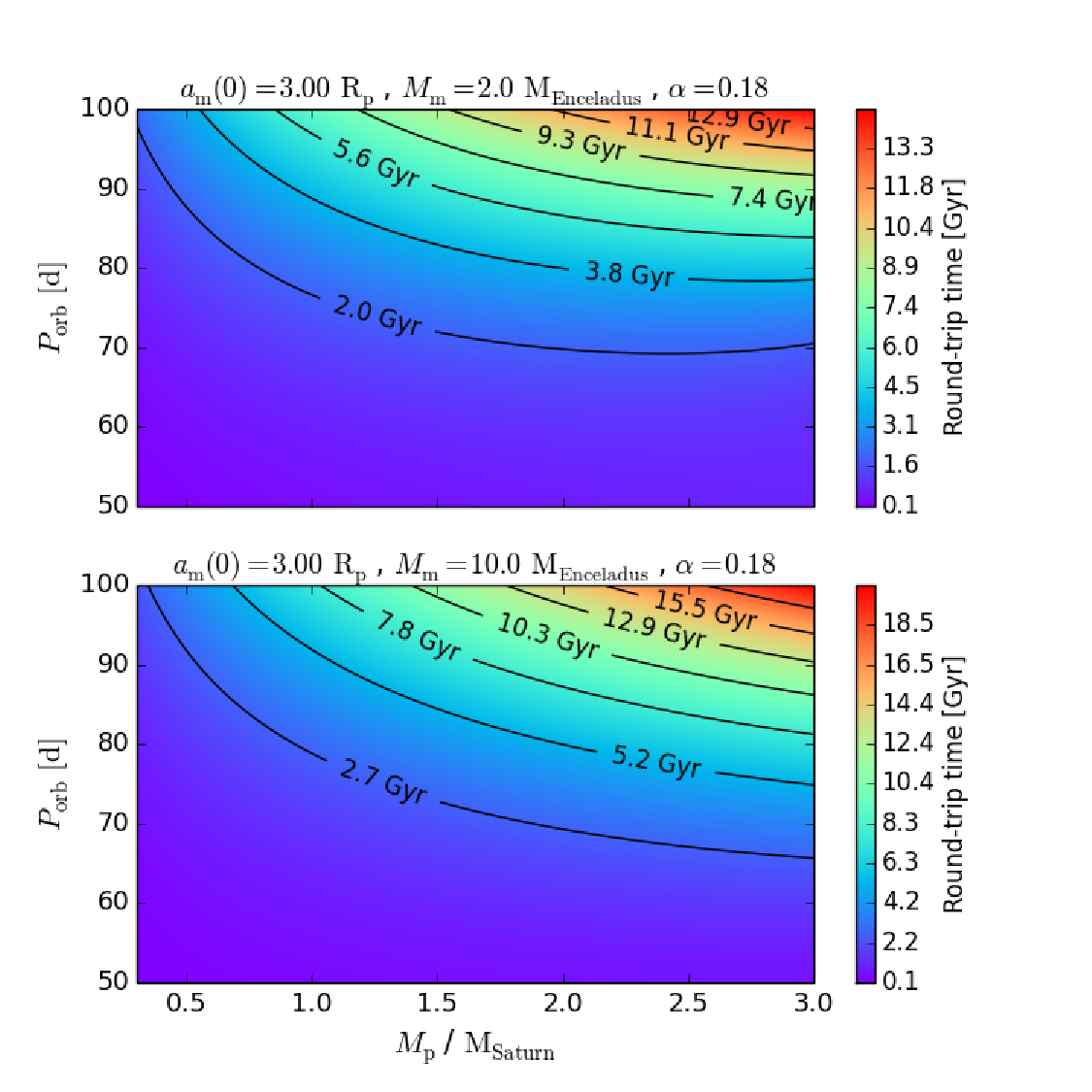}
\caption{Contour plots of the total round-trip time-scale $\trt$ by assuming different properties for the planet-moon system. In all of the cases we consider $\Mstar=1\,\Msun$.}
\label{fig:quasistaticTimescales}
\end{figure*}

\hl{For illustration purposes, in \autoref{fig:quasistaticEvolution} we show the results of integrating equations (\ref{eq:dOdt}) and (\ref{eq:dnmdt}) for a particular case: a Saturn-mass planet orbiting a solar mass star with $P\sub{orb}=60-80$ days ($a\sub{p}\approx 0.3-0.4$ au) and having a moon 10 times the mass of Enceladus at $\amoon(0)=3\Rp$.  Hereafter, we will use this `warm Saturn-Superenceladus' system to illustrate the results in each scenario and perform comparison among them.  It should be understood, however, that the conclusions of our comparison are not substantially modified, if we use a system with different properties.  Only the moon migration time-scales will change.  Precisely, and in order to gain some insight on the dependence of the relevant time-scales of the properties in the system, we show in \autoref{fig:quasistaticTimescales} contour plots of round-trip times in the \quasistatic\ scenario.}

\hl{Migration time-scales are very sensitive to planetary distance (see right column in \autoref{fig:quasistaticTimescales}).  Since stellar torque on the planetary bulge goes as $P\sub{orb}^4$ (Eq. \autoref{eq:Tors}), closer planets loss their rotational angular momentum faster, so that moons do not reach distant positions and synchronization between $\Op$ and $\nmm$ is achieved much earlier.}

\hl{Time-scales are also very different among systems having different planetary and moon masses. More massive planets have a larger rotational inertia and hence larger synchronization times.  On the other hand more massive moons have larger orbital angular momentum and outwards/inwards migration takes also longer (see left column in \autoref{fig:quasistaticTimescales}).}

\subsection{Unresponsive migration}

\hl{The first significant effect of planetary evolution on moon migration is observed when we let that $\Rp$ and $\kappa^2$ change in time.   To disentangle the effect of planetary evolution from other effects at play, we will assume that, despite the obvious effect that a varying $\Rp$ will have in the tidal response of the planet, tidal dissipation reservoir $\Kpp/\Qp$ remains constant.  We call this scenario \unresponsive\ migration.}

\begin{figure}
 \centering
 \includegraphics[scale=0.462]{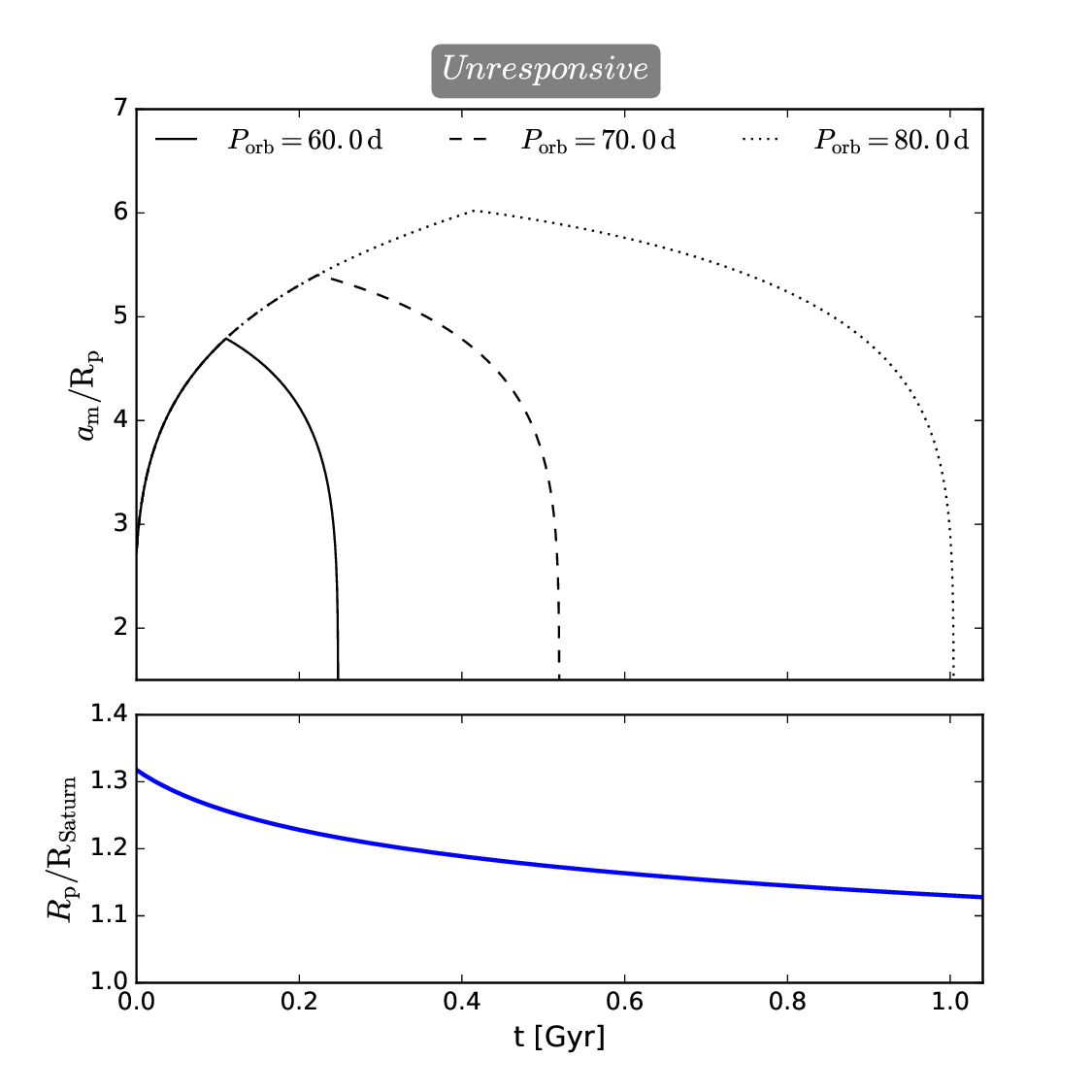}
 \caption{Upper plot: moon orbital semi major axis as function of time for a warm Saturn-Superenceladus system in the \unresponsive\ scenario.  Lower plot: evolution of planetary radius. Solid, dashed and dotted lines correspond to planetary orbital periods of 60, 70 and 80 days, respectively.}
 \label{fig:unresponsiveEvolution}
 \end{figure}

\hl{In \autoref{fig:unresponsiveEvolution} we show the result of integrating Eqs. (\ref{eq:dOdt},\ref{eq:dnmdt}) for the same case studied in the \quasistatic\ case, but including now a variable planetary radius (lower plot).}

\begin{figure}
\centering
\includegraphics[scale=0.462]{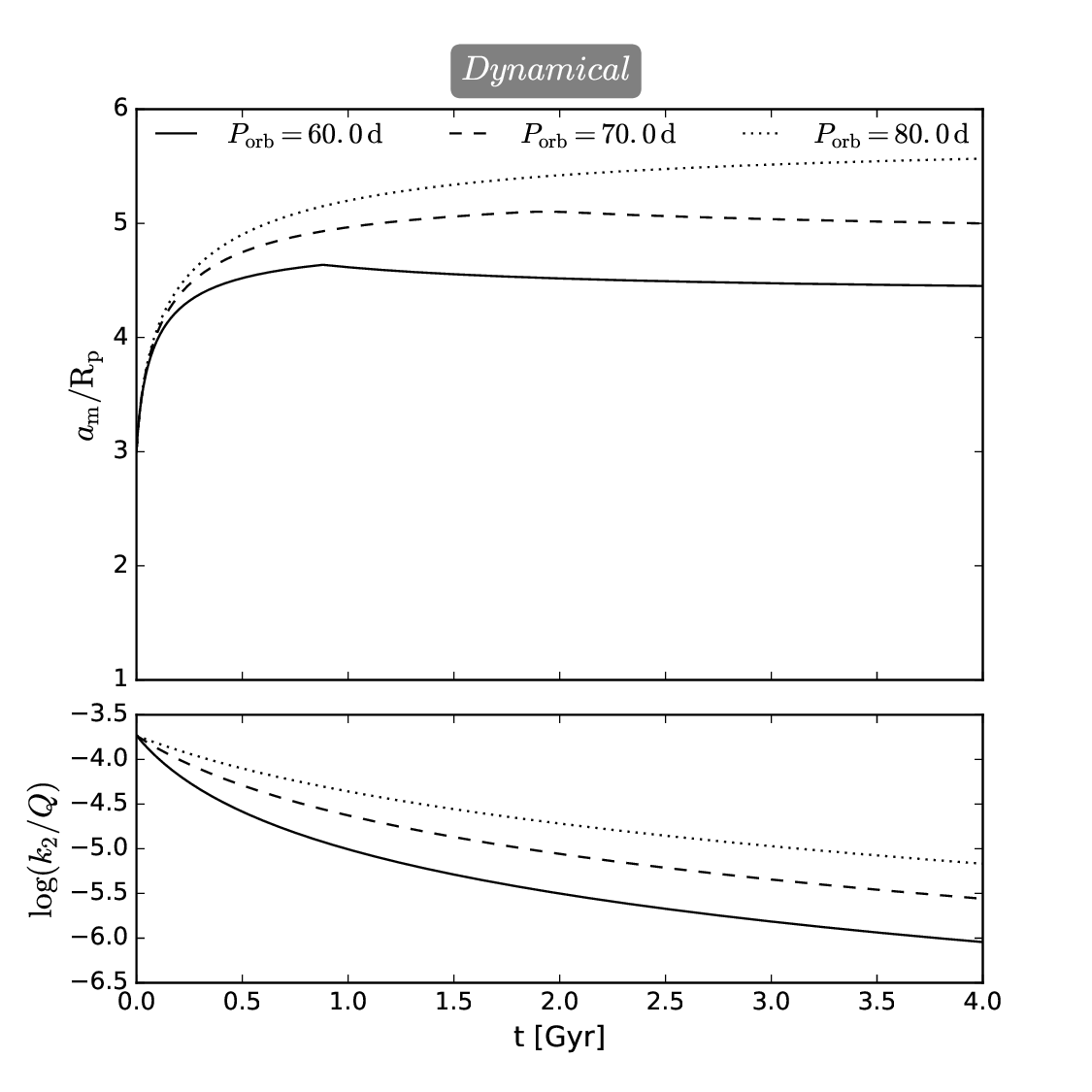}
\caption{Upper plot: moon orbital semi major axis as function of time for a warm Saturn-Superenceladus system. Lower plot: evolution of $\Kpp/\Qp$. Solid, dashed and dotted lines correspond to planetary orbital periods of 60, 70 and 80 days, respectively. }
\label{fig:dynamicalEvolution}
\end{figure}

\hl{The main effect that the evolution of planetary radius has on moon orbital migration, arises from the fact that an evolving young planet is effectively larger than a static one.  Since tidal torques scale-up with radius as $\Rp^5$, even a $\sim 30$ per cent larger radius, produce tidal torques $\sim 4$ times larger at earlier phases of planetary evolution as compared to those produced in the \quasistatic\ case.  This effect speeds up both moon orbital migration and planetary rotational \hll{braking}. As a result, migration timescales are reduced by almost a factor of 2 (see \autoref{fig:unresponsiveEvolution}).}

\subsection{Dynamical migration}

\hl{One of the key features of our model is the consistent calculation of the tidal dissipation on giant planets as a function of their evolving properties.  As we have already shown in Section \ref{sec:migration}, turbulent friction of inertial waves sustained by Coriolis forces is the main driver of tidal dissipation in fluid giant planets. As such, it will strongly depend on the rotational velocity of the planet, which varies as the planet interact with the star and the moon.}

\hl{Our third evolutionary scenario, the \dynamic\ scenario, assumes that the radius of the planet $\Rp$ and its interior structure remain almost constant whereas the rotational rate $\Op$ changes in time.  In \autoref{fig:dynamicalEvolution} we show the effect that a varying $\Op$ has on the tidal dissipation reservoir $\Kpp/\Qp$ and the impact that those changes have on the moon orbital evolution.} 

\hl{At the beginning, when the planet is rotating faster, $\Kpp/\Qp$ has its largest value. In this phase moon migration and planetary rotational \hll{braking} occur in a shorter time-scale than in the \quasistatic\ case. As the planet breaks, $\Kpp/\Qp$ falls by almost one order of magnitude in several hundreds of Myr. Less tidal energy is dissipated by inertial waves in the liquid envelope of the planet, and moon migration stalls.}

\hl{We have verified that independent of moon initial distance to the planet, planet or moon mass, or planetary distance to its host star, moon migration around a liquid giant planet will never end in a course of collision or disruption of the moon inside the Roche radius.  This result is in starking contrast with what was previously expected \citep{Barnes2002,Sasaki2012}}. \hll{ However, it is worth to note that final conclusions can be affected according to the dissipation mechanism adopted in the process, which can lead to different scenarios from those proposed in this work. For instance, the apparently strange behaviour of Saturn's moons' tidal evolution implies that Saturn's internal tidal dissipation may be strongly peaked in forcing frequency space, leading to nonlinear effects that are not considered for our purposes.}

\subsection{Realistic migration}

\begin{figure}
 \centering
 \includegraphics[scale=0.468]{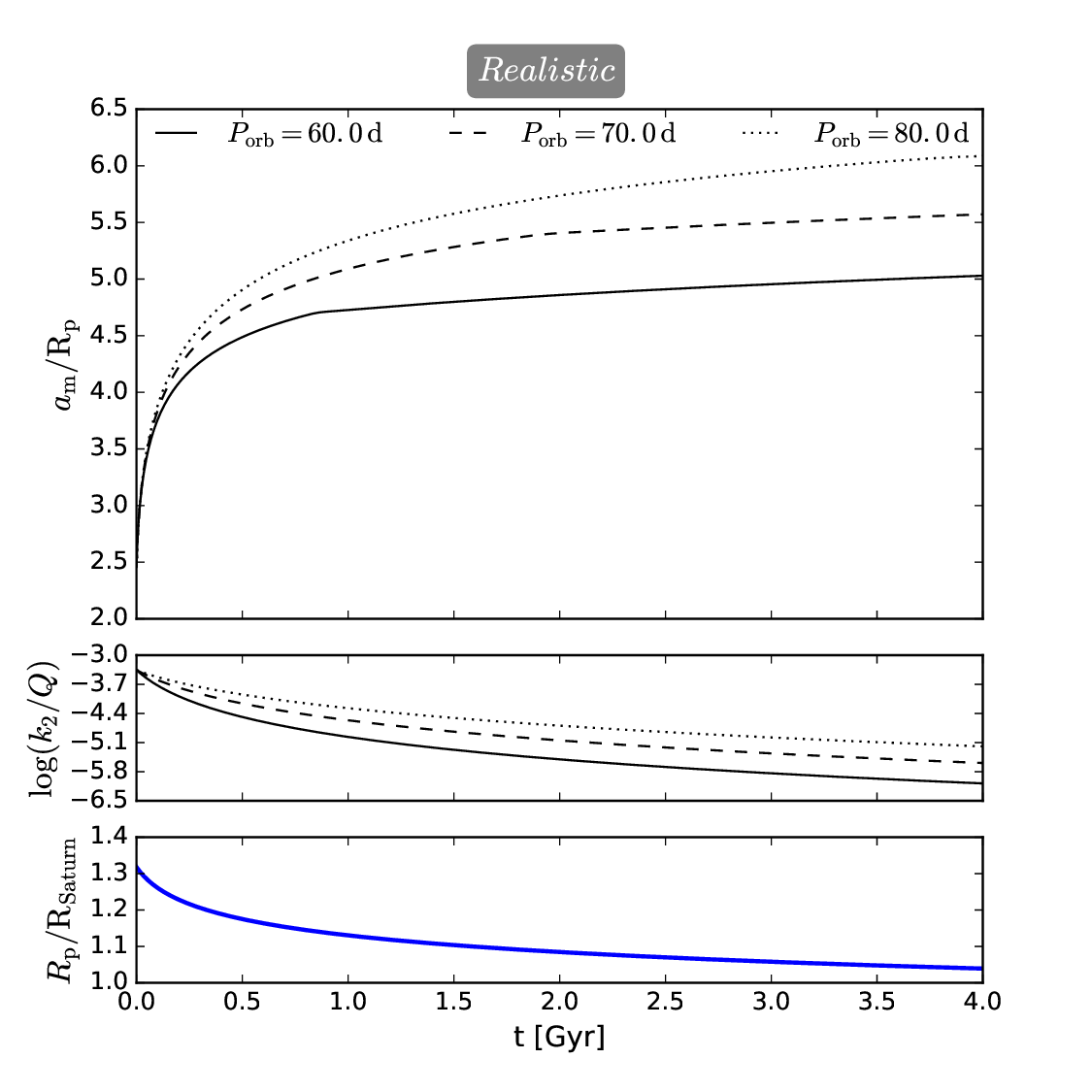}
 \caption{Upper plot: moon orbital semi major axis as function of time for a warm Saturn-Superenceladus system in the \unresponsive\ scenario.  Middle plot: evolution of $\Kpp/\Qp$. Lower plot: evolution of planetary radius. Solid, dashed and dotted lines correspond to planetary orbital periods of 60, 70 and 80 days, respectively.}
 \label{fig:realisticEvolution}
 \end{figure}

Finally\hl{, the \realistic\ scenario is that on which all} the properties (i.e. $\Rp$\hl{, $\kappa$, $\Op$} and $\Kpp/\Qp$) \hl{ vary as functions of time}. \hl{The result of including all the effects on moon orbital evolution is shown in} \autoref{fig:realisticEvolution}. 

\hl{As in the \dynamic\ case, tidal torques are large at the beginning, accelerating moon migration and planetary rotational \hll{braking}.  Once the moon has reached the synchronous distance it seems to stall.  However, since the planet is still contracting, and the moon is even migrating inwards, its distance, as measured in planetary radii seems to be monotonically increasing.  Although the increase in the relative distance does not modify the strength of the gravitational or tidal forces on the moon, it contributes to monotonically drive the moon away of a collision or its tidal disruption.} In large time-scales this would cause that moons pass along the critical semi-major axis at $\sim0.48\,\RH$ \citep{Domingos2006} becoming their orbits unstable and prone to be ejected from planetocentric to heliocentric orbits. Therefore, this situation leads to the possibility that under certain circumstances, the moon becomes a new dwarf rocky planet, something that we call a `ploonet'.

It is noticeable that \dynamic\, and \realistic\, situations seem to have a similar behaviour. Even so, the first one exhibits an inflection point at short time-scales which is not prominent in the second one. Hence, the \dynamic\, scenario tends to park the moon in a quasi-stationary orbit, constraining any future orbital evolution. On the contrary, the \realistic\, case allows the moon to migrate outward indefinitely. In other words, the spin-orbit synchronization is always present for short time-scales in situations where the planet's size remains constant over time. The difference between both cases lies in that the \realistic\, scenario couples the gradual change of $\Rp$ into the evolution of $\Kpp/\Qp$, by means of equation (\ref{eq:k2QFormula}), influencing the underlying dynamic and therefore the final fate of the moon.

\section{Summary and Conclusions}
\label{sec:discon}

\hl{In this paper we have used the tidal model of \citet{Barnes2002}, recently extended and updated by \citet{Sasaki2012}, to study the orbital evolution of exomoons around close-in giant planets. As a novel feature we have consistently included in those models, the evolution of the physical bulk properties of the planet and its evolving response to tidal stresses.}

\hl{Our ``evolutionary'' moon migration model, relies on the well-known results of close-in giant planet thermal evolution by \citet{Fortney2007}, which predict the bulk properties of planets with different composition and at different distances from their host-star.  To model the response of the planet to tidal stresses and predict tidal dissipation of rotational energy and transfer of angular momentum towards the moon, we use the recent analytical models of \citet{Ogilvie2013}.  In that model the value of the frequency averaged ratio $\Kpp/\Op$ is calculated as a function of planetary bulk properties and rotational rate.}

\hl{For our numerical experiments, and in particular for the calculation of $\Kpp/\Op$, we assume that the interior of giant-planets can be modeled as constituted by two constant density layers, an outer one, made of a low density liquid, and a central liquid/solid core.  We assumed that the dissipation of tidal energy occurs via turbulent friction of inertial waves in the liquid outer layer.  We neglected the contribution of tidal dissipation inside the core.}

\hl{We found that orbital migration of exomoons is significantly modified when the evolution of planetary bulk properties is included.  In the most studied case where the planet bulk properties evolve very slowly with respect to moon migration time-scales, and tidal dissipation is completely independent of planetary rotational rate (\quasistatic  \ scenario), moons migrate outwards and then inwards, facing a collision with the planet or tidal obliteration/disruption in relatively short time-scales (several Gyrs).  If only the planetary radius changes (\unresponsive \ scenario), outwards then inwards migration is faster and the fate of the moons are similar.  However if we let the tidal response to depend on rotational rate (as is the case if tidal energy is dissipated in the liquid outer envelope of the planet) and/or if the planet contracts in a similar or shorter time-scale (i.e. \dynamic\,and \realistic\, scenarios, respectively), the moon never falls back into the planet. This result is independent of the planet, star and moon masses or their mutual distance.}

\hl{This is a completely unexpected, still fortunate, outcome of the effect of planetary evolution on moon orbital migration. If confirmed by further more detailed models, large and detectable regular exomoons may have survived orbital migration around already discovered exoplanets and could be awaiting a future detection.}

\hl{Interestingly, the dependence of an exomoon's final fate on the evolution of their host close-in giant planets, could be used to constraint planetary evolutionary models, and/or to model the planet's interior structure and its response to tidal stresses.  In the forthcoming future, when missions such as Transiting Exoplanet Survey Satellite (TESS), Characterizing Exoplanets Satellite (CHEOPS), the James Webb Space Telescope (JWST) and Planetary Transits and Oscillations of stars (PLATO) (see \citealt{Rauer2014} and references therein) hopefully provide us a definitive confirmation (or rejection) of the existence of exomoons around close-in planets, these results will be confirmed.}


\section*{Acknowledgements}
We thank the referee J. W. Barnes, whose valuable comments allowed us to improve the manuscript. We also appreciate the useful comments to the initial versions of this manuscript provided by R. Canup. J.A.A. is supported by the Young Researchers program of the Vicerrector\'ia de Investigaci\'on, J.I.Z. by Vicerrector\'ia de Docencia of the Universidad de Antioquia (UdeA) and M.S. by Doctoral Program of Colciencias and the CODI/UdeA. This work is supported by Vicerrectoria de Docencia-UdeA and the {\it Estrategia de Sostenibilidad 2016-2017 de la Universidad de Antioquia}.  Special thanks to Nadia Silva for the review of the manuscript.



\bsp	
\label{lastpage}
\end{document}